\documentclass[reprint, aps]{revtex4-2}
\usepackage{amsfonts}
\usepackage{amssymb}
\usepackage{amsmath}
\usepackage{graphics}
\usepackage{graphicx}
\usepackage{subfigure}
\usepackage{dcolumn}
\usepackage{bm}
\usepackage{hyperref}
\usepackage{color}
\usepackage{epstopdf}

\DeclareMathAlphabet\mathbfcal{OMS}{cmsy}{b}{n}

\begin{document}

\title{High-order harmonic generation in three-dimensional Weyl semimetals}
\author{H. K. Avetissian}
\author{V. N. Avetisyan}
\author{ B. R. Avchyan}
\author{G. F. Mkrtchian}
\thanks{mkrtchian@ysu.am}

\affiliation{Centre of Strong Fields Physics, Yerevan State University,
Yerevan 0025, Armenia}

\begin{abstract}
In this paper, the nonlinear interaction of Weyl semimetal (WSM) with a
strong driving electromagnetic wave-field is investigated. In the scope of
the structure-gauge invariant low-energy nonlinear electrodynamic theory,
the polarization-resolved high-order harmonic generation spectra in WSM are
analyzed. The obtained results show that the spectra in WSM are completely
different compared to 2D graphene case. In particular, at the non-collinear
arrangement of the electric and Weyl nodes' momentum separation vectors, an
anomalous harmonics are generated which are polarised perpendicular to the
pump wave electric field. The intensities of anomalous harmonics are
quadratically dependent on the momentum space separation of the Weyl nodes.
If the right and the left Weyl fermions are merged, we have a 4-component
trivial massless Dirac fermion and, as a consequence, the anomalous
harmonics vanish. In contrast to the anomalous harmonics, the intensities of
normal harmonics do not depend on the Weyl nodes' momentum separation
vector, and the harmonics spectra resemble the picture for a massless 3D
Dirac fermion.
\end{abstract}

\maketitle

\section{ Introduction}

As a three-dimensional analogs of graphene \cite{grph1,grph2}, the Dirac
semimetals (DSM) \cite{dsm1,dsm2,dsm3,dsm4} and the WSM \cite%
{wsm1,wsm2,wsm3,wsm4,wsm5,wsm6} have been implemented in a variety of
condensed matter systems. These materials are three-dimensional quantum
phases of the matter with gapless electronic excitations that are protected
by topology and symmetry \cite{Armitage}. The low energy dispersion of such
materials contains conical intersections and diabolical points, which are
referred to as a Dirac points, or a Weyl nodes \cite{Wan}. The DSMs possess
both time-reversal and spatial inversion symmetry. When one of these
symmetries is broken, the Dirac points are split into the pair of the Weyl
nodes, and the medium becomes a WSM. The low energy theory of the simplest
WSM is described by the Weyl Hamiltonian \cite{Weyl} near the Weyl nodes
where the right-handed and the left-handed chirality fermions are separated
in the momentum space. Due to the nontrivial topology of the bands, the
Berry curvature in the momentum space is nonzero \cite{Berry,Xiao}, and we
have an appropriate case of the Dirac monopole/anti-monopole \cite{Dirac}
realization in the momentum space \cite{Fang}. As a result, the linear
electromagnetic (EM) response of the three-dimensional WSM is described by
an axionic field theory \cite{Adler,Bell,Wilczek} with the anomalous linear
electrodynamic effects \cite{Zyuzin,Son,Wang1,Goswami,Vazifeh}. The
interaction between the strong EM waves and WSM gives rise to nonlinear
optical effects, such as the photovoltaic effect \cite{Osterhoudt,Ma},
optical rectification and second-harmonic generation \cite{Wu,Wang,Takasan},
terahertz emission \cite{Gao} and third harmonic generation \cite{Tilmann}.
These are perturbative nonlinear optical effects. With the further increase
of the driving wave intensity, the extreme nonlinear optical effects \cite%
{Avetissian-book} may be visible in pseudo-relativistic systems. In
particular, the high-order harmonic generation (HHG) is an essential
nonlinear dynamic process that can be used as a probe to extract the
properties of a medium. It can also be useful for new nanodevices. To date,
HHG has been observed in graphene \cite{Yoshikawa}, in DSM \cite{Lim,Kovalev}%
, in topological insulators \cite{Bai}, and in WSM \cite{Lv}, where the
\textquotedblleft spike-like\textquotedblright\ Berry curvature may generate
even-order harmonics. Note that as in graphene, there is quite a high
carrier mobility in WSM \cite{Shekhar,Kumar}, that is the electrons can move
significantly in the Brillouin zone, which is favorable for HHG phenomenon
in nanostructures.

As is well studied for graphene the HHG process at Dirac-cone approximation 
\cite{Mikh,Mer,H4,Al-Naib,H7,Mer18} significantly different from the HHG
when electrons can move significantly in the Brillouin zone, here
polarization and optical anisotropy effects of HHG in graphene arise \cite%
{Zurr,Liu,H12,H14,Wang3, H15,H13,Feng,Mer2022a}. The HHG in WSM with
particular lattice realization theoretically is studied in Ref. \cite{Dixit}
where anisotropic anomalous HHG from time-reversal symmetry broken WSM is
reported. Nonperturbative topological intraband current in WSM and DSM in
laser fields has been investigated in Ref. \cite{Dantas} for general case,
without lattice concretization.\ To establish a nonlinear response
intrinsically connected to topology one should arise from the universal Weyl
Hamiltonian which is the root of the field theory anomalies. Hence, there is
tremendous interest from the strong fields physics perspectives in
understanding how the field theory anomalies affect the nonlinear response
of WSM at low energy excitations where the theory is universal and does not
depend on the particular lattice realization of the WSM. To this end, in the
current paper we investigate the low-energy nonlinear electrodynamics of WSM
and analyze polarization-resolved high-order harmonic generation spectra in
WSM. The consideration is based on structure-gauge invariant low-energy
nonlinear electrodynamics where an ansatz applied to the Dirac monopole \cite%
{Topology} is adopted to overcome the topological singularity.

The paper is organized as follows. In Sec. II the structure-gauge invariant
low-energy nonlinear electrodynamic theory with evolutionary equation for
the single-particle density matrix is presented. In Sec. III, we consider
polarization-resolved HHG spectra and present the main results. Finally,
conclusions are given in Sec. IV.

\section{Model Hamiltonian and the evolutionary equation for the
single-particle density matrix}

We will start from the low energy universal Hamiltonian involving
4-component massless Dirac fermion:%
\begin{equation}
\widehat{H}_{0}=\mathrm{v}\int d^{3}x\bar{\Psi}\left[ -i\gamma ^{j}\partial
_{j}+b_{\mu }\gamma ^{\mu }\gamma ^{5}\right] \Psi  \label{h1}
\end{equation}%
where $\mathrm{v}$ is the Fermi velocity, $\bar{\Psi}=\Psi ^{\dagger }\gamma
_{0}$, matrices ${\gamma ^{0}}$ and ${\gamma ^{j}}$ ($j=1,2,3$) are Dirac
anticommuting $\gamma $ matrices, $\gamma ^{5}=i\gamma ^{0}\gamma ^{1}\gamma
^{2}\gamma ^{3}$ is the chirality matrix, and $b_{\mu }$ is the axial
4-vector. In the chiral representation of the $\gamma $ matrices we have 
\begin{equation}
{\gamma ^{0}}=\left( 
\begin{array}{cc}
0 & \mathit{1} \\ 
\mathit{1} & 0%
\end{array}%
\right) ,\;\boldsymbol{\gamma }=\left( 
\begin{array}{cc}
0 & \widehat{\boldsymbol{\sigma }} \\ 
{-}\widehat{\boldsymbol{\sigma }} & 0%
\end{array}%
\right) ,\;{\gamma ^{5}}=\left( 
\begin{array}{cc}
-\mathit{1} & 0 \\ 
0 & \mathit{1}%
\end{array}%
\right) ,\;  \label{gamma}
\end{equation}%
where $\widehat{\boldsymbol{\sigma }}$ is the vector operator formed out of
three Pauli matrices. In Eq. (\ref{h1}) the term proportional to $b_{\mu }$
breaks $CPT$ symmetry. In this paper we will consider the case of space-like
axial vector $b_{0}=0$. In this case axial vector term $\mathbf{b\gamma }$
preserves inversion ($\mathcal{P}$) and charge conjugation ($\mathcal{C}$)
symmetries, but breaks time reversal symmetry ($\mathcal{T}$). Note that
this case is more feasible for the realization of a Weyl semi-metal and the
corresponding minimal lattice model can easily be constructed \cite{Vazifeh}%
. The Hamiltonian (\ref{h1}) in the momentum space becomes%
\begin{equation}
\widehat{H}_{0}\left( \mathbf{k}\right) =\mathrm{v}\left( 
\begin{array}{cc}
\mathbf{\sigma \cdot }\left( \mathbf{k+b}\right) & 0 \\ 
{0} & -\mathbf{\sigma \cdot }\left( \mathbf{k-b}\right)%
\end{array}%
\right) .  \label{h2}
\end{equation}%
For compactness of equations atomic units are used throughout the paper
unless otherwise indicated. The eigenstates of this Hamiltonian are also
eigenstates of chirality matrix ${\gamma ^{5}}$ with eigenvalues $\chi =\pm
1 ${. Since Dirac mass is zero, }the $\widehat{H}_{0}\left( \mathbf{k}%
\right) $ is block diagonal and {the }left-handed $\left( 1-{\gamma ^{5}}%
\right) \Psi /2$ and right-handed components $\left( 1+{\gamma ^{5}}\right)
\Psi /2$ of the Dirac field are decoupled to left-handed and right-handed
two-component Weyl spinors, described by the Hamiltonians 
\begin{equation}
\widehat{H}_{\chi }=-\chi \mathrm{v}\mathbf{\sigma \cdot }\left( \mathbf{k-}%
\chi \mathbf{b}\right) ;\chi =\pm 1.  \label{h3}
\end{equation}

The $2\times 2$ Hamiltonians $\widehat{H}_{1}$ and $\widehat{H}_{-1}$ also
describe the monopole and the anti-monopole of the Berry curvature in the
momentum space, respectively \cite{Xiao,Topology}. For $\mathbf{b}\neq 0$,
the right and the left Weyl fermions are separated in the momentum space and
the WSM is topologically non-trivial. The eigenvalues $\chi =\pm 1$ also
have topological notion. The Berry flux piercing any surface enclosing the
Weyl nodes $\mathbf{k=}\chi \mathbf{b}$ is exactly $2\pi \chi $, i.e. $\chi $
also defines the Chern number or topological charge. In accordance with
Nielsen-Ninomiya theorem the Weyl nodes should come in opposite chirality
pairs \cite{NN}. When $\mathbf{b}=0$ these Weyl nodes are merged giving rise
to topologically trivial, 4-component massless Dirac fermion.

For the calculation of the nonlinear EM response of WSM we need eigenstates
of the Hamiltonian (\ref{h3}). With these eigenstates we should calculate
Berry connection and then curvature. Because of the monopole in momentum
space, the eigenstates of the Weyl Hamiltonian (\ref{h3}) cannot be defined
globally for all $\mathbf{k}$. The eigenstates $|\beta ,\chi ,\mathbf{k}%
\rangle $, where $\beta $ refers to band index, can be subject to an
arbitrary structure-gauge \cite{Yue}\textit{\ }transformation%
\begin{equation}
|\beta ,\chi ,\mathbf{k}\rangle ^{\prime }=e^{i\vartheta _{\chi \beta
}\left( \mathbf{k}\right) }|\beta ,\chi ,\mathbf{k}\rangle   \label{eigen}
\end{equation}%
without changing the physical properties of the system. For the quantum
kinetics we need to calculate the transition dipole moments $\mathbf{d}%
_{\beta \beta ^{\prime }}\left( \chi ,\mathbf{k}\right) =\langle \beta ,\chi
,\mathbf{k}|i\partial _{\mathbf{k}}|\beta ^{\prime },\chi ,\mathbf{k}\rangle 
$. The Berry connections are defined as the diagonal elements $\mathbfcal{A}%
_{\beta }\left( \chi ,\mathbf{k}\right) =\mathbf{d}_{\beta \beta }\left(
\chi ,\mathbf{k}\right) $. Hence, due to the gradient $\partial _{\mathbf{k}}
$ a smooth structure gauge for the eigenstates is thus required. To overcome
this problem we will adopt the ansatz applied to Dirac monopole. To this end
we choose axial vector $\mathbf{b}$ directed along the x-axis $\mathbf{b=}b%
\widehat{\mathbf{x}}$ and define the eigenstates for $k_{z}\geq 0$ and $%
k_{z}\leq 0$. Since $\widehat{H}_{\chi =1}\left( -\mathbf{k}\right) =%
\widehat{H}_{\chi =-1}\left( \mathbf{k}\right) $ we will bring the solution
for the left-handed Weyl spinors. The eigenstates $|\beta ,\chi ,\mathbf{k}%
\rangle _{+}$ for $k_{z}\geq 0$ are:%
\begin{equation}
|c,\chi ,\mathbf{k}\rangle _{+}=\frac{1}{\sqrt{2k_{\chi }^{2}+2k_{\chi }k_{z}%
}}\left[ 
\begin{array}{c}
k_{\chi }+k_{z} \\ 
k_{x\chi }+ik_{y}%
\end{array}%
\right] ,  \label{cp}
\end{equation}%
\begin{equation}
|v,\chi ,\mathbf{k}\rangle _{+}=\frac{1}{\sqrt{2k_{\chi }^{2}+2k_{\chi }k_{z}%
}}\left[ 
\begin{array}{c}
-k_{x\chi }+ik_{y} \\ 
k_{\chi }+k_{z}%
\end{array}%
\right] ,  \label{vp}
\end{equation}%
for $k_{z}\leq 0$ we have%
\begin{equation}
|c,\chi ,\mathbf{k}\rangle _{-}=\frac{1}{\sqrt{2k_{\chi }^{2}-2k_{\chi }k_{z}%
}}\left[ 
\begin{array}{c}
k_{x\chi }-ik_{y} \\ 
k_{\chi }-k_{z}%
\end{array}%
\right] ,  \label{cm}
\end{equation}%
\begin{equation}
|v,\chi ,\mathbf{k}\rangle _{-}=\frac{1}{\sqrt{2k_{\chi }^{2}-2k_{\chi }k_{z}%
}}\left[ 
\begin{array}{c}
-k_{\chi }+k_{z} \\ 
k_{x\chi }+ik_{y}%
\end{array}%
\right] ,  \label{vm}
\end{equation}%
where $k_{x\chi }=k_{x}-\chi b$ and $k_{\chi }=\sqrt{k_{x\chi
}^{2}+k_{y}^{2}+k_{z}^{2}}$. The eigenenergies are: $\mathcal{E}_{c\chi
}\left( \mathbf{k}\right) =\mathrm{v}k_{\chi }$ and $\mathcal{E}_{v\chi
}\left( \mathbf{k}\right) =-\mathrm{v}k_{\chi }$.

The solutions for the opposite chirality are:\textrm{\ }$|c,\chi ,\mathbf{k}%
\rangle _{\pm }=|c,-\chi ,-\mathbf{k}\rangle _{\mp }$ and \textrm{\ }$%
|v,\chi ,\mathbf{k}\rangle _{\pm }=|v,-\chi ,-\mathbf{k}\rangle _{\mp }$. At
the overlap $k_{z}=0$ solutions (\ref{cp}), (\ref{cm}), and (\ref{vp}), (\ref%
{vm}) are connected by the gauge transformation:%
\begin{eqnarray*}
|c,\chi ,\mathbf{k}\rangle _{-} &=&e^{-i\vartheta \left( k_{x\chi
},k_{y}\right) }|c,\chi ,\mathbf{k}\rangle _{+}, \\
|v,\chi ,\mathbf{k}\rangle _{-} &=&e^{i\vartheta \left( k_{x\chi
},k_{y}\right) }|v,\chi ,\mathbf{k}\rangle _{+},
\end{eqnarray*}%
where $\vartheta \left( k_{x\chi },k_{y}\right) =\arctan \left(
k_{y}/k_{x\chi }\right) $. From Eqs. (\ref{cp}), (\ref{cm}), (\ref{vp}), and
(\ref{vm}) for the total Berry connection $\mathbfcal{A}\left( \chi ,\mathbf{k}%
\right) =\langle c,\chi ,\mathbf{k}|i\partial _{\mathbf{k}}|c,\chi ,\mathbf{k%
}\rangle -\langle v,\chi ,\mathbf{k}|i\partial _{\mathbf{k}}|v,\chi ,\mathbf{%
k}\rangle $ we obtain: 
\begin{equation}
\mathcal{A}_{\mathrm{+}}\left( \chi ,\mathbf{k}\right) =-\chi \frac{k_{y}%
\widehat{\mathbf{x}}-k_{x\chi }\widehat{\mathbf{y}}}{k_{\chi }^{2}+k_{\chi
}k_{z}},  \label{Ap}
\end{equation}%
\begin{equation}
\mathcal{A}_{\mathrm{-}}\left( \chi ;\mathbf{k}\right) =\chi \frac{k_{y}%
\widehat{\mathbf{x}}-k_{x\chi }\widehat{\mathbf{y}}}{k_{\chi }^{2}-k_{\chi
}k_{z}}.  \label{Am}
\end{equation}%
For the Berry curvature $\mathbfcal{B}\left( \chi ;\mathbf{k}\right) =\partial 
\mathbf{k\times }\mathcal{A}\left( \chi ,\mathbf{k}\right) $ we obtain
located at the Weyl node $\mathbf{k=}\chi \mathbf{b}$ monopole field%
\begin{equation}
\mathcal{B}_{x}=\chi \frac{k_{x\chi }}{k_{\chi }^{3}},\ \mathcal{B}_{y}=\chi 
\frac{k_{y}}{k_{\chi }^{3}},\ \mathcal{B}_{z}=\chi \frac{k_{z}}{k_{\chi }^{3}%
},  \label{monopole}
\end{equation}%
with $\mathrm{div}\mathbfcal{B}=4\pi \chi \delta \left( \mathbf{k-}\chi 
\mathbf{b}\right) $ and $\delta $ is the Dirac delta function. For the
transition dipole moments we have

\begin{equation*}
\mathbf{d}_{cv_{+}}\left( \chi ,\mathbf{k}\right) =\frac{\left( ik_{x\chi
}-\chi k_{y}\right) \widehat{\mathbf{z}}}{2k_{\chi }^{2}}+\frac{1}{2k_{\chi
}\left( k_{\chi }+k_{z}\right) }
\end{equation*}%
\begin{equation*}
\times \left[ \left( i\left( \frac{k_{x\chi }^{2}}{k_{\chi }}-k_{\chi
}-k_{z}\right) -\frac{\chi k_{y}k_{x\chi }}{k_{\chi }}\right) \widehat{%
\mathbf{x}}\right. 
\end{equation*}%
\begin{equation}
\left. +\left( \chi \left( k_{\chi }+k_{z}-\frac{k_{y}^{2}}{k_{\chi }}%
\right) +i\frac{k_{x\chi }k_{y}}{k_{\chi }}\right) \widehat{\mathbf{y}}%
\right] ,  \label{tdmp}
\end{equation}%
\begin{equation*}
\mathbf{d}_{cv_{-}}\left( \chi ,\mathbf{k}\right) =\frac{\left( ik_{x\chi
}+\chi k_{y}\right) \widehat{\mathbf{z}}}{2k_{\chi }^{2}}-\frac{1}{2k_{\chi
}\left( k_{\chi }-k_{z}\right) }
\end{equation*}%
\begin{equation*}
\times \left[ \left( \chi \frac{k_{y}k_{x\chi }}{k_{\chi }}+i\left( \frac{%
k_{x\chi }^{2}}{k_{\chi }}-k_{\chi }+k_{z}\right) \right) \widehat{\mathbf{x}%
}\right. 
\end{equation*}%
\begin{equation}
\left. -\left( \chi \left( k_{\chi }-k_{z}-\frac{k_{y}^{2}}{k_{\chi }}%
\right) -\frac{ik_{x\chi }k_{y}}{k_{\chi }}\right) \widehat{\mathbf{y}}%
\right] .  \label{tdmm}
\end{equation}%
Note the following useful relations 
\begin{equation*}
\frac{1}{2}\epsilon ^{abc}\mathcal{B}^{c}\left( \chi ,\mathbf{k}\right) 
\end{equation*}%
\begin{equation}
=i\left\{ d_{cv}^{a}\left( \chi ,\mathbf{k}\right) d_{vc}^{b}\left( \chi ,%
\mathbf{k}\right) -d_{cv}^{b}\left( \chi ,\mathbf{k}\right) d_{vc}^{a}\left(
\chi ,\mathbf{k}\right) \right\} ,  \label{db1}
\end{equation}%
where $\epsilon ^{abc}$ is the Levi-Civita symbol and the summation over the
repeated upper indices is implied. This equation is gauge invariant and
connects the transition dipole moments with the Berry curvature. Here for
the sake of brevity, we omit the indices ($\pm $).

The semiconductor Bloch equations (SBEs) governing a WSM driven by a strong
laser field in the length gauge read:%
\begin{equation*}
\partial _{t}\rho _{\alpha \beta ;\chi }(\mathbf{k}_{0},t)=i\mathcal{E}%
_{\beta \alpha ;\chi }\left( \mathbf{k}_{0}+\mathbf{A}\right) \rho _{\alpha
\beta ;\chi }(\mathbf{k}_{0},t)
\end{equation*}%
\begin{equation*}
-\left( 1-\delta _{\alpha \beta }\right) \Gamma \rho _{\alpha \beta ;\chi }(%
\mathbf{k}_{0},t)+i\left[ \sum\limits_{\alpha ^{\prime }}\mathbf{d}_{\alpha
^{\prime }\beta }\left( \chi ,\mathbf{k}_{0}\mathbf{+A}\right) \rho _{\alpha
\alpha ^{\prime };\chi }(\mathbf{k}_{0},t)\right.
\end{equation*}%
\begin{equation}
\left. -\sum\limits_{\beta ^{\prime }}\mathbf{d}_{\alpha \beta ^{\prime
}}\left( \chi ,\mathbf{k}_{0}\mathbf{+A}\right) \rho _{\beta ^{\prime }\beta
;\chi }(\mathbf{k}_{0},t)\right] \mathbf{E}(t),  \label{SBE}
\end{equation}%
where $\rho _{\alpha \beta ;\chi }$ are the single particle density matrix
elements, $\mathbf{E}$ is the laser electric field strength, $\mathbf{A=-}%
\int_{0}^{t}\mathbf{E}\left( t^{\prime }\right) dt^{\prime }$ is the vector
potential, $\mathcal{E}_{\beta \alpha ;\chi }\left( \mathbf{k}\right) =%
\mathcal{E}_{\beta \chi }\left( \mathbf{k}\right) -\mathcal{E}_{\alpha \chi
}\left( \mathbf{k}\right) $, and $\Gamma ^{-1}$ is the dephasing time. The
crystal momentum $\mathbf{k}$ has been transformed into a frame moving with
the vector potential $\mathbf{k}_{0}=\mathbf{k}-\mathbf{A}$. Note that in
Eq. (\ref{SBE}), the Berry connections (\ref{Ap}) and (\ref{Am}) are
included $\mathbf{d}_{\beta \beta }\left( \chi ,\mathbf{k}\right) \widehat{=}%
\mathbf{A}_{\beta }\left( \chi ,\mathbf{k}\right) $.

The optical excitation induces a volume current that can be calculated by
the following formula:%
\begin{equation*}
\mathbf{j}\left( t\right) =-\left[ \sum\limits_{\alpha \chi \mathbf{k}%
_{0}}\left( \mathbf{V}_{\alpha \chi }\left( \mathbf{k}_{0}\mathbf{+A}\right)
\right) \rho _{\alpha \alpha ;\chi }(\mathbf{k}_{0},t)\right.
\end{equation*}
\begin{equation}
\left. +i\sum\limits_{\alpha \neq \beta ,\chi }\sum\limits_{\mathbf{k}_{0}}%
\mathbf{d}_{\beta \alpha }\left( \chi ,\mathbf{k}_{0}\mathbf{+A}\right) 
\mathcal{E}_{\beta \alpha ;\chi }\left( \mathbf{k}_{0}\mathbf{+A}\right)
\rho _{\alpha \beta ;\chi }(\mathbf{k}_{0},t)\right] ,  \label{vcur}
\end{equation}%
where the band velocity is defined by $\mathbf{V}_{\alpha \chi }\left( 
\mathbf{k}\right) =\partial _{\mathbf{k}}\mathcal{E}_{\alpha \chi }(\mathbf{%
k)}$. Note that Eqs. (\ref{SBE}) and (\ref{vcur}) provide structure-gauge%
\textit{\ }invariant kinetic theory. Thus, at the structure-gauge
transformation (\ref{eigen}) we have%
\begin{equation*}
\mathbf{d}_{\alpha \beta }^{\prime }\left( \chi ,\mathbf{k}\right)
=e^{i\theta _{\chi \beta }\left( \mathbf{k}\right) -i\theta _{\chi \alpha
}\left( \mathbf{k}\right) }\mathbf{d}_{\alpha \beta }\left( \chi ,\mathbf{k}%
\right) ,
\end{equation*}%
\begin{equation*}
\mathbfcal{A}_{\alpha }^{\prime }\left( \chi ,\mathbf{k}\right) =\mathbfcal{A}%
_{\alpha }\left( \chi ,\mathbf{k}\right) -\partial _{\mathbf{k}}\theta
_{\chi \alpha }\left( \mathbf{k}\right) ,\mathbfcal{B}^{\prime }\left( \chi ;%
\mathbf{k}\right) =\mathbfcal{B}\left( \chi ;\mathbf{k}\right) ,
\end{equation*}%
\begin{equation*}
\rho _{\alpha \beta ;\chi }^{\prime }(\mathbf{k}_{0},t)=e^{i\theta _{\beta
}\left( \mathbf{k}_{0}\mathbf{+A}\right) -i\theta _{\alpha }\left( \mathbf{k}%
_{0}\mathbf{+A}\right) }\rho _{\alpha \beta ;\chi }(\mathbf{k}_{0},t),
\end{equation*}%
\begin{equation*}
\mathbf{j}^{\prime }\left( t\right) =\mathbf{j}\left( t\right) .
\end{equation*}


\section{Results}

We explore the nonlinear response of a WSM in a laser field of ultrashort
duration:%
\begin{equation}
\mathbf{E}\left( t\right) =f\left( t\right) E_{0}\hat{\mathbf{e}}\cos \left(
\omega t\right) ,  \label{field}
\end{equation}%
where $f\left( t\right) =\sin ^{2}\left( \pi t/\tau \right) $ is the
sin-squared envelope function, $\tau $ is the pulse duration, $\hat{\mathbf{e%
}}$ is the unit polarization vector, $\omega $ is the currier frequency, $%
E_{0}$ is the electric field amplitude. We take a ten-cycle fundamental
laser field. 
\begin{figure*}[tbp]
\includegraphics[width=.99\textwidth]{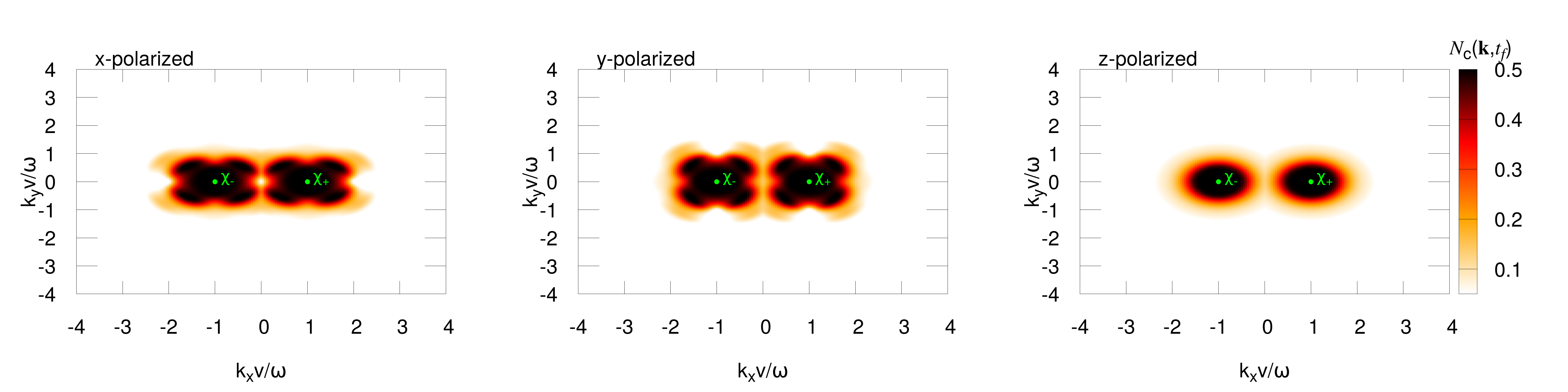}
\caption{Particle distribution function $\mathcal{N}\left( \mathbf{k}%
,t_{f}\right) $ (in arbitrary units) in the plane $k_{z}=0$ after the
interaction at the instant $t_{f}=\protect\tau $ for WSM, as a function of
scaled dimensionless momentum components ($k_{x}\mathrm{v}/\protect\omega $, 
$k_{y}\mathrm{v}/\protect\omega $) for different orientations of the laser
electric field strength. The wave-particle dimensionless interaction
parameter is taken to be $\protect\xi _{0}=0.5$, and the axial vector
magnitude is chosen to be $b=\protect\omega /\mathrm{v.}$ The Weyl nodes are
located at $k_{x}\mathrm{v}/\protect\omega =\pm 1$\textrm{.}}
\end{figure*}

As in graphene, the wave-particle interaction in WSM is characterized by the
dimensionless parameter \cite{Mer}%
\begin{equation}
\xi _{0}=\frac{eE_{0}\mathrm{v}}{\omega }\frac{1}{\hbar \omega },
\label{par}
\end{equation}%
which represents the work of the wave electric field $E_{0}$ on a period $%
1/\omega $ in the units of photon energy $\hbar \omega $. \ The parameter is
written here in general units for clarity. For two band WSM system SBEs (\ref%
{SBE}) are reduced to a closed set of equations for the interband
polarization $P_{\chi }(\mathbf{k}_{0},t)\equiv \rho _{vc;\chi }(\mathbf{k}%
_{0},t)$ and for the distribution functions $N_{c/v}\left( \mathbf{k}%
_{0},t\right) \equiv \rho _{cc/vv;\chi }(\mathbf{k}_{0},t)$ in the
conduction/valence bands. For an undopped system in equilibrium, the initial
conditions $P_{\chi }(\mathbf{k}_{0},0)=0$, $N_{c}(\mathbf{k}_{0},0)=0$, and 
$N_{v}(\mathbf{k}_{0},0)=1$ are assumed, neglecting thermal occupations. The
integration of SBEs is performed on a 3D grid of $500\times 500\times 500$
points homogeneously distributed in the cube $\left( -\alpha _{cut}\omega /%
\mathrm{v},\alpha _{cut}\omega /\mathrm{v}\right) _{XYZ}$. The
minimum/maximum crystal momentum is defined by $\alpha _{cut}$, which in
turn depends on the intensity of the pump wave. The time integration is
performed with the standard fourth-order Runge-Kutta algorithm. From Eq. (%
\ref{vcur}) follows the relation:%
\begin{equation}
\frac{d\mathbf{j}\left( t\right) }{dt}\frac{\mathrm{v}^{2}}{\omega ^{4}}=%
\mathbf{w}\left(\overline{t},\xi _{0}, \frac{b\mathrm{v}}{\omega}, \frac{\Gamma }{\omega }\right) ,
\label{diml}
\end{equation}%
where $\overline{t}=\omega t$, and $\mathbf{w}\left(\overline{t},\xi _{0}, 
\frac{b\mathrm{v}}{\omega}, \frac{\Gamma }{\omega }\right) $ is a periodic (in case of an external
monochromatic wave) dimensionless universal function that parametrically
depends on the WSM--wave interaction parameters $\xi _{0}$, the scaled axial vector, and the scaled
relaxation rate. Hence, by solving SBEs (\ref{SBE}), performing the integral
over $\mathbf{k}_{0}$ (\ref{vcur}) and taking Fourier transform ($\mathcal{FT%
}$), the polarization- resolved high-harmonic spectrum is calculated as 
\begin{equation}
I_{\alpha }=\left\vert \mathcal{FT}\left( w_{\alpha }\left( \overline{t},\xi _{0}, 
\frac{b\mathrm{v}}{\omega}, \frac{\Gamma }{\omega }\right) \right) \right\vert ^{2},\ \alpha
=x,y,x.  \label{int}
\end{equation}%
For all calculations, the relaxation time is taken to be equal to half of
the wave period $\Gamma ^{-1}=T/2=\pi /\omega $. 

The typical photoexcitation of the Fermi-Dirac sea is presented in Fig. 1,
where the density plot of the particle distribution function $N_{c}\left( 
\mathbf{k},t_{f}\right) $ after the interaction at the instant $t_{f}=10T$ ,
as a function of dimensionless momentum components, for different
orientations of the laser electric field strength are shown. As is seen from
this figure, near the Weyl nodes we have an almost homogeneous excitation
due to the singularity of the transition dipole moments. Far from the Weyl
nodes the excitation pattern is defined by the anisotropy of the transition
dipole moments (\ref{tdmp}) and (\ref{tdmm}). 
\begin{figure}[tbp]
\includegraphics[width=.45\textwidth]{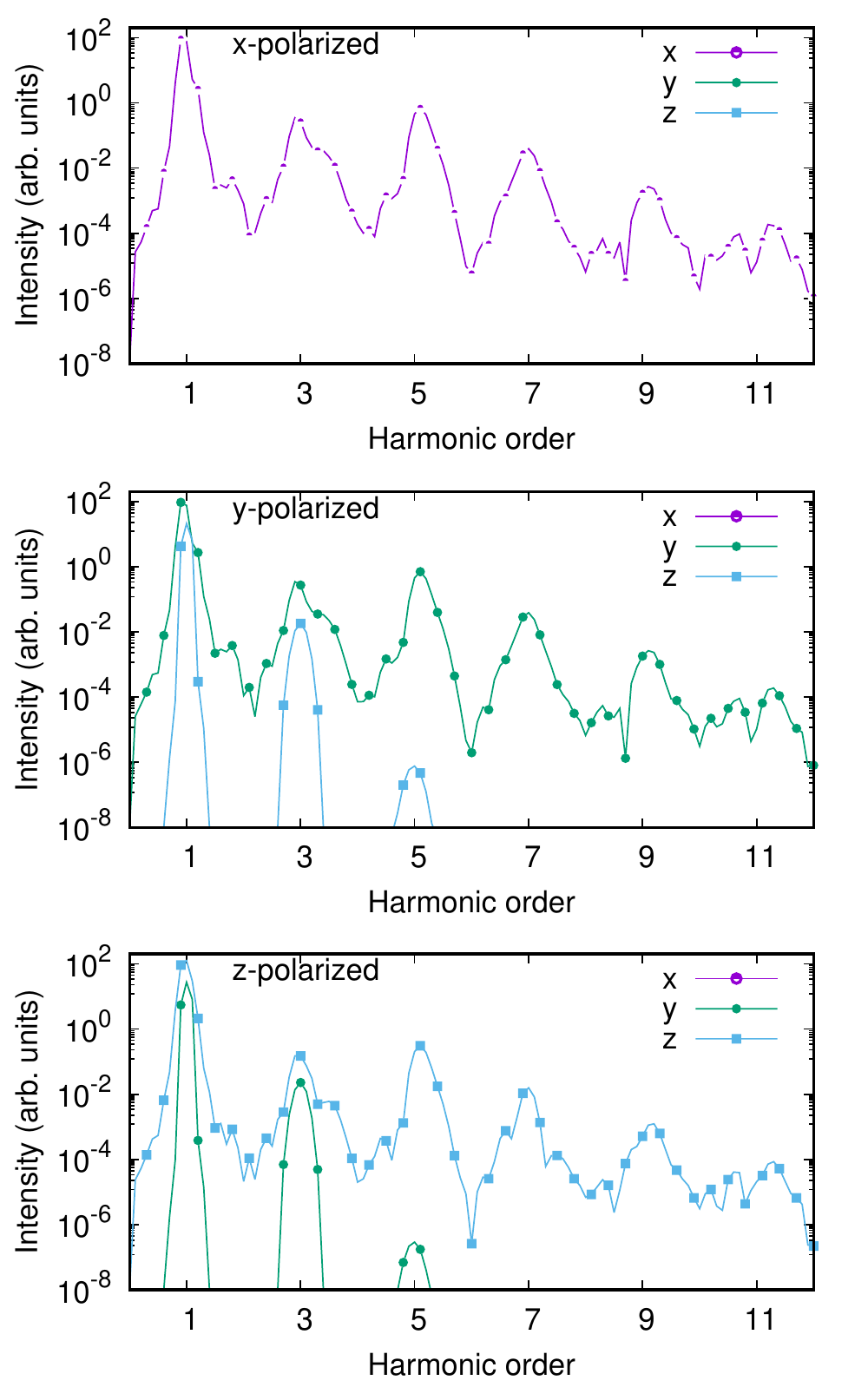}
\caption{The polarization resolved HHG spectra in logarithmic scale for WSM
in the strong-field regime for different orientations of the laser electric
field strength. The wave-particle dimensionless interaction parameter is
taken to be $\protect\xi _{0}=0.5$, and the axial vector magnitude is
choosen to be $b=\protect\omega /2\mathrm{v.}$ The Weyl nodes are located at 
$k_{x}\mathrm{v}/\protect\omega =\pm 0.5.$}
\end{figure}
\begin{figure}[tbp]
\includegraphics[width=.45\textwidth]{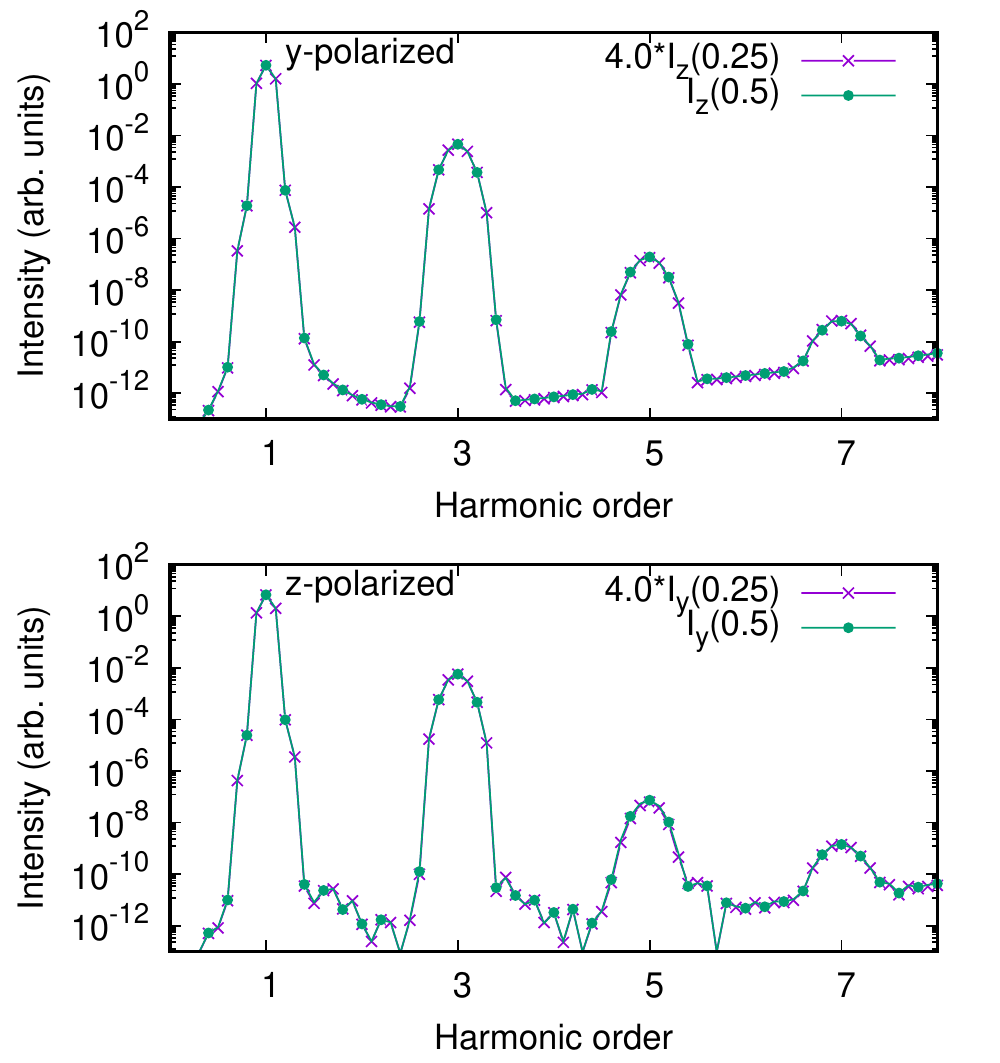}
\caption{The anomalous HHG spectra in logarithmic scale for WSM in the
strong-field regime for different axial vector magnitudes: $b\mathrm{v}/%
\protect\omega =0.25$ and $b\mathrm{v}/\protect\omega =0.5$. The
wave-particle dimensionless interaction parameter is taken to be $\protect%
\xi _{0}=1.0$.}
\end{figure}

In Fig. 2, the polarization-resolved HHG spectra in logarithmic scale for
WSM in the strong-field regime for different orientations of the laser
electric field strength are presented. From top to bottom we show the
spectra for the $x$, $y$, and $z$ polarizations of the pump wave. As is seen
from this figure, when the driving wave is polarized along $x$ direction,
the odd harmonics are generated only along the laser polarization direction.
However, when the wave is polarized along $y$ or $z$ directions, in addition
to normal harmonics generated along the laser polarization, anomalous
harmonics along perpendicular directions are also generated. As reflected
from Fig. 2, anomalous harmonics are generated at the non-collinear
arrangement of the electric field and Weyl node's momentum separation
vectors. This is the manifestation of the axionic field theory with the
anomalous nonlinear electrodynamic effects. To understand how these findings
are related to the non-trivial topology of WSM, let us derive another
equivalent equation for the current (\ref{vcur}) that explicitly includes
Berry curvature (\ref{monopole}). From Eq. (\ref{SBE}), inserting expression
for $i\mathcal{E}_{\beta \alpha \chi }\left( \mathbf{k}_{0}+\mathbf{A}%
\right) \rho _{\alpha \beta ;\chi }(\mathbf{k}_{0},t)$ into the equation for
the current (\ref{vcur}), taking into account the relation (\ref{db1}),
electron-hole symmetry $\mathbf{V}_{v\chi }=-\mathbf{V}_{c\chi }$, and the
integral of motion $N_{v;\chi }(\mathbf{k}_{0},t)+N_{c;\chi }(\mathbf{k}%
_{0},t)=1$, we find%
\begin{equation*}
j_{a}\left( t\right) =-2\sum\limits_{\chi \mathbf{k}_{0}}V_{c\chi
}^{a}N_{c;\chi }(\mathbf{k}_{0},t)
\end{equation*}%
\begin{equation*}
-2\mathrm{Re}\sum\limits_{\chi \mathbf{k}_{0}}d_{cv}^{a}\left( \chi ,\mathbf{%
k}_{0}\mathbf{+A}\right) \left\{ \partial _{t}P_{\chi }(\mathbf{k}%
_{0},t)+\Gamma P_{\chi }(\mathbf{k}_{0},t)\right\} 
\end{equation*}%
\begin{equation*}
+2\mathrm{Re}\sum\limits_{\chi \mathbf{k}_{0}}iE^{b}(t)\mathcal{A}^{b}\left(
\chi ,\mathbf{k}_{0}\mathbf{+A}\right) d_{cv}^{a}\left( \chi ,\mathbf{k}_{0}%
\mathbf{+A}\right) P_{\chi }(\mathbf{k}_{0},t)
\end{equation*}%
\begin{equation*}
+\sum\limits_{\chi \mathbf{k}_{0}}\frac{1}{2}\epsilon ^{abc}E^{b}(t)\mathcal{%
B}^{c}\left( \chi ,\mathbf{k}_{0}\mathbf{+A}\right) 
\end{equation*}%
\begin{equation}
-\sum\limits_{\chi \mathbf{k}_{0}}\epsilon ^{abc}E^{b}(t)\mathcal{B}%
^{c}\left( \chi ,\mathbf{k}_{0}\mathbf{+A}\right) N_{c;\chi }(\mathbf{k}%
_{0},t),  \label{111}
\end{equation}%
where the summation over the repeated upper indices is implied. The first
term in Eq. (\ref{111}) is the ordinary intraband part of the current, the
second and third terms define the interband part of the current, the fourth
and fifth terms represent the topological part of the current. These terms
are defined by the Berry curvature (\ref{monopole}). The fourth term is
nothing but the anomalous Hall current \cite{Goswami,Vazifeh} and depends
linearly on the field strength $E$, since $\sum\limits_{\chi \mathbf{k}_{0}}%
\mathcal{B}^{c}\left( \chi ,\mathbf{k}_{0}\mathbf{+A}\right)
=\sum\limits_{\chi \mathbf{k}}\mathcal{B}^{c}\left( \chi ,\mathbf{k}\right) $%
. The last term in Eq. (\ref{111}) is the nonlinear part of the anomalous
Hall current that gives rise to anomalous harmonics perpendicular to the
laser field strength directions: $\epsilon ^{abc}E^{b}\mathcal{B}^{c}=%
\mathbf{E}\times\mathbfcal{B}$. \ If the driving wave is polarized along the
axial-vector $\mathbf{b}$ ($x$-direction), then it is easy to see that the
anomalous current along $y$ and $z$ directions turn out to be zero as
monopole fields (\ref{monopole}) of Weyl nodes cancel each other. On the
other hand, when the driving wave is polarized along the $y$ or $z$
directions, then the $x$-component of the Berry curvature comes into play.
Let us analyze for the concreteness the case of the $z$-polarized driving
wave. In this case the anomalous Hall current can be written%
\begin{equation*}
j_{y}=\frac{E^{z}(t)}{(2\pi )^{3}}\int d^{3}\mathbf{k}_{0}
\end{equation*}%
\begin{equation}
\times \left[ \frac{\left( k_{0x}+b\right) N_{c;1}(\mathbf{k}_{0},t)}{\sqrt{%
\left( k_{0x}+b\right) ^{2}+k_{0y}^{2}+\left( k_{0z}+A_{z}\right) ^{2}}}%
-(b\rightarrow -b)\right] .  \label{mi}
\end{equation}%
At the first glance for each Weyl node taking into account unbounded linear
dispersion of fermions one can make a naive shift of variable $k_{x\chi
}=k_{x}-\chi b$, and make this integral to vanish. However, we should take
into account singularity near $\pm b$ points as in the case of linear
axionic field theory \cite{Goswami}. Therefore we need to choose a finite
cut-off along the axial vector ($x$-direction), which can be sent to
infinity at the end of calculations and can keep the cut-offs in the
directions perpendicular to the axial vector to be infinity. As reflected
from Fig. 1, near the Weyl nodes we have an almost homogeneous excitation
due to the singularity of the transition dipole moments (\ref{tdmp}) and (%
\ref{tdmm}). Hence, we can approximate the integral (\ref{mi}) as%
\begin{equation*}
j_{y}\simeq \frac{2}{(2\pi )^{3}}E^{z}(t)N_{c;1}(\mathbf{k}_{0w},t)
\end{equation*}%
\begin{equation*}
\times \int_{0}^{\infty }k_{\perp }dk_{\perp }\int_{-\Lambda }^{\Lambda
}dk_{0x}\int_{0}^{2\pi }d\phi \frac{k_{0x}+b}{\left[ k_{\perp
}^{2}+(k_{0x}+b)^{2}\right] ^{3/2}}
\end{equation*}%
\begin{equation*}
=\frac{2}{(2\pi )^{2}}E^{z}(t)N_{c;1}(\mathbf{k}_{0w},t)\int_{-\Lambda
}^{\Lambda }\mathrm{sgn}(k_{0x}+b)dk_{0x}
\end{equation*}%
\begin{equation}
=\frac{b}{\pi ^{2}}E^{z}(t)N_{c;1}(\mathbf{k}_{0w},t),  \label{mi2}
\end{equation}%
\begin{figure}[tbp]
\includegraphics[width=.45\textwidth]{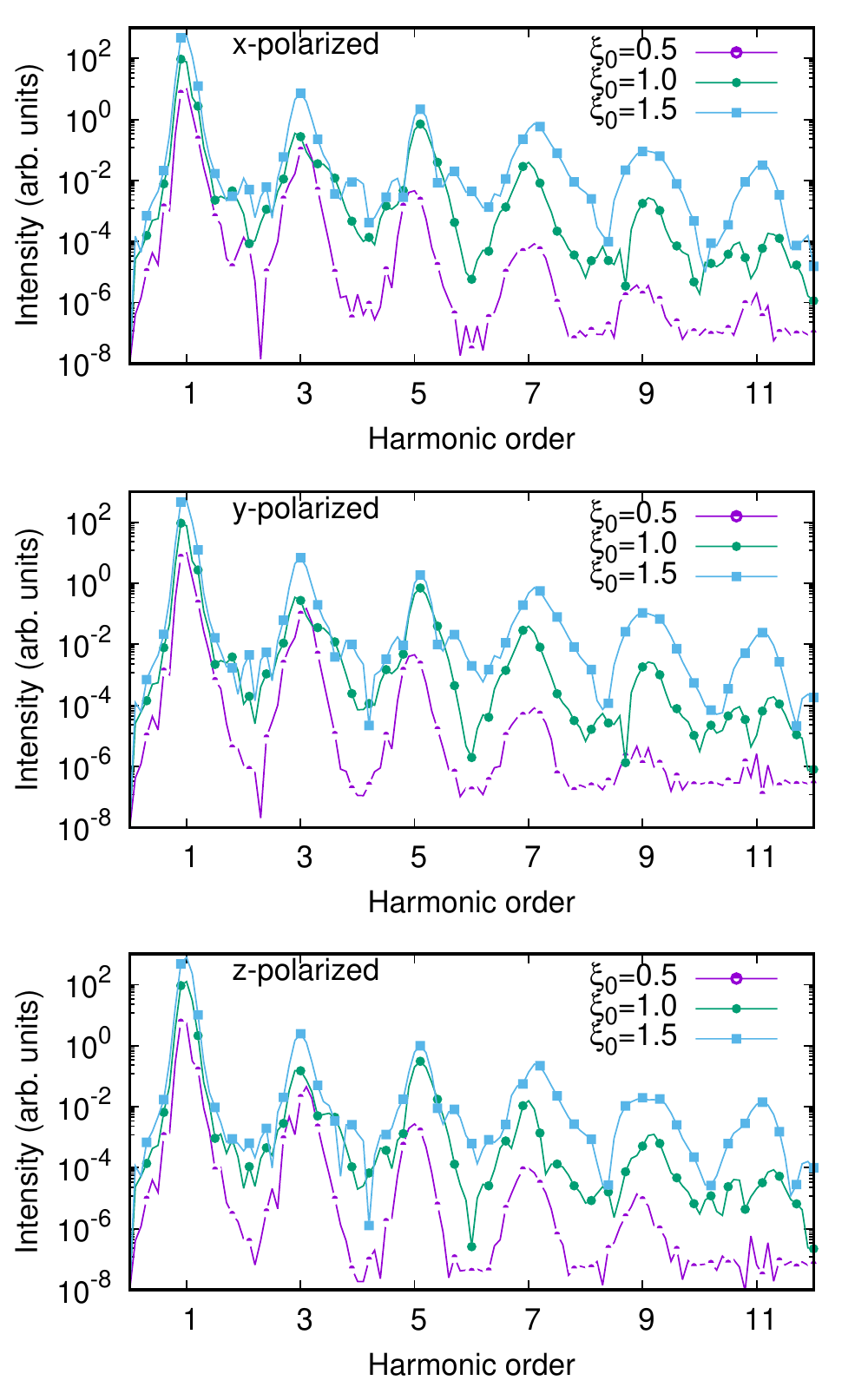}
\caption{The HHG spectra for the normal harmonics in logarithmic scale for
WSM in the strong-field regime for different orientations of the pump laser
electric field strength at various wave-particle dimensionless interaction
parameter $\protect\xi _{0}$.}
\end{figure}
where $N_{c;1}(\mathbf{k}_{0w},t)$ is calculated near the Weyl node: $%
\mathbf{k}_{0w}=\left( b,0,0\right) $. This is an interesting result that
implies that the nonlinear anomalous Hall current is proportional to
axial-vector as in the linear axionic field theory. But, this result is only
valid for the unbounded linear dispersion of the Weyl fermions. If we
consider a lattice model of WSM, the linear dispersion is valid only near
the Weyl nodes, and integration is performed over the finite Brillouin zone.
Therefore, the nonlinear anomalous Hall current of WSM at high energies will
be modified by a non-linear contribution from the axial-vector. In Fig. 3,
the anomalous HHG spectra in the logarithmic scale for WSM in the
strong-field regime, for different axial vector magnitudes are presented.
The results for $b\mathrm{v}/\omega =0.25$ are multiplied by the factor 4.
As is seen from Fig. 3, the intensities of anomalous harmonics are
quadratically dependent on the momentum space separation of the Weyl nodes.
This is consistent with our approximate (\ref{mi2}) result. This differs
from the case of the lattice model \cite{Dixit} where the intensity of
anomalous harmonics decreases with the increasing distance between the Weyl
nodes. It is straightforward to see that for the normal harmonics with the
shift of variable $k_{x\chi }=k_{x}-\chi b$ one can obtain the results
independent on $b$. This is equivalent to the fact that the right and the
left Weyl fermions are merged, we have a 4-component trivial massless Dirac
fermion and, as a consequence, the anomalous harmonics vanish. In addition,
the intensities of normal harmonics do not depend on the Weyl node's
momentum separation vector and resemble the results for a massless 3D Dirac
fermion.

We now turn to an examination of the effect of the driving wave intensity on
the HHG in WSM. We present the results of simulations for normal harmonics
at different polarizations in Fig. 4. The intensities of normal harmonics $%
I_{\alpha }$ do not depend on the Weyl node's location. For the considered
intensities the perturbation theory is not applicable, and in Fig. 4 we have
a strong deviation from the power law for the intensities of harmonics. In
particular, the intensities of the 5th, 7th, and 9th harmonics scale as $%
I_{5}\sim \xi _{0}^{3}$, $I_{7}\sim \xi _{0}^{7}$, and $I_{9}\sim \xi
_{0}^{9}$, respectively. Whereas they should show the $I_{n}\sim \xi
_{0}^{2n}$ dependence in the perturbative limit. Besides, this figure shows
that the intensities of the normal harmonics are almost independent of the
pump wave polarization, which is connected with the isotropic linear
dispersion of the Weyl fermions.

In Fig. 5, the HHG spectra for the anomalous harmonics for WSM in the
strong-field regime for different orientations of the pump laser electric
field strength at various wave-particle dimensionless interaction parameter $%
\xi _{0}$ are shown. The Weyl nodes are located at $k_{x}\mathrm{v}/\omega
=\pm 0.5$. In this case, we also have a strong deviation from the power law
for the intensities of anomalous harmonics. In particular, the intensities
of the 3rd, 5th, and 7th harmonics scale as $I_{3}\sim \xi _{0}^{4}$, $%
I_{5}\sim \xi _{0}^{6}$, and $I_{7}\sim \xi _{0}^{8}$, respectively. The
dependences of the intensities of normal and anomalous harmonics on the
intensity of the driving wave are completely different, which is due to the
different underlying mechanisms.

\begin{figure}[tbp]
\includegraphics[width=.45\textwidth]{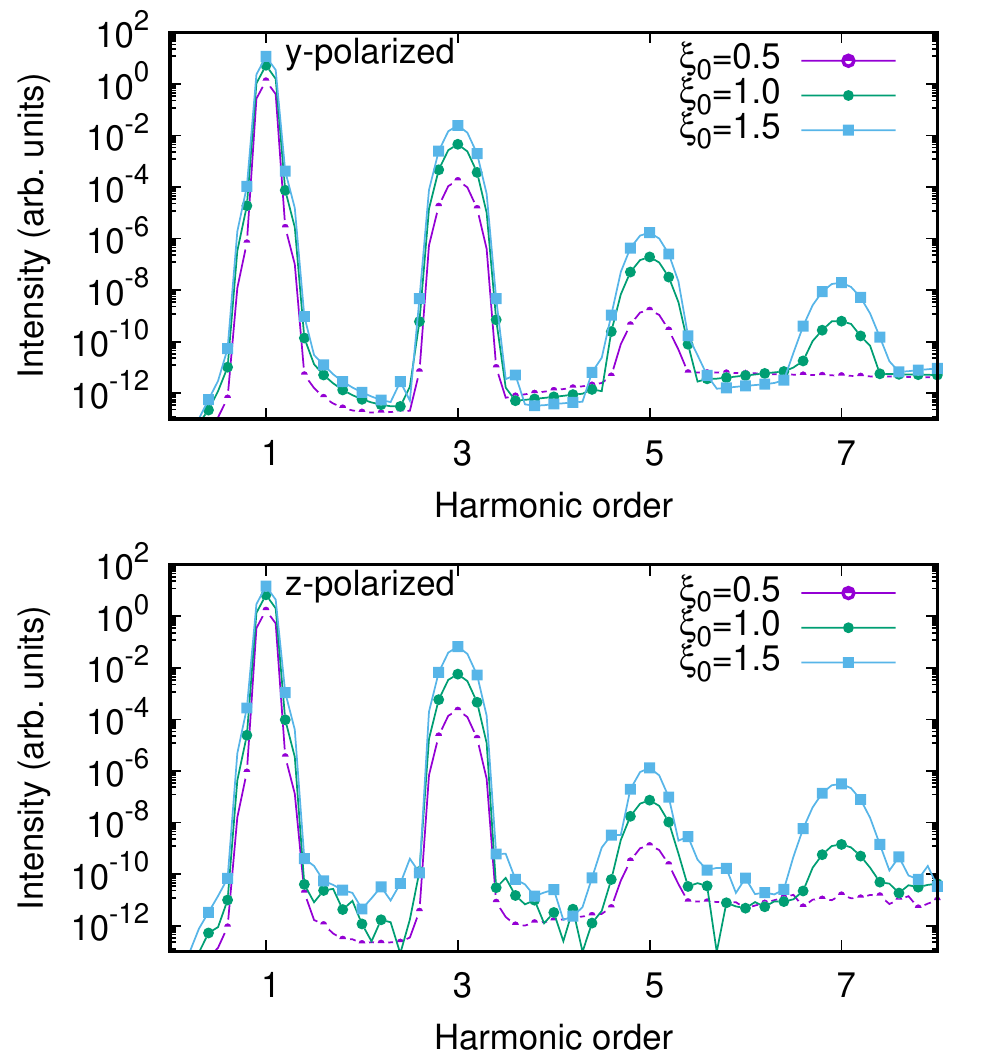}
\caption{The HHG spectra for the anomalous harmonics in logarithmic scale
for WSM in the strong-field regime for different orientations of the pump
laser electric field strength at various wave-particle dimensionless
interaction parameter $\protect\xi _{0}$. The Weyl nodes are located at $%
k_{x}\mathrm{v}/\protect\omega =\pm 0.5.$}
\end{figure}

\section{Conclusion}

We have presented the structure-gauge invariant microscopic theory of
nonlinear interaction of a time-reversal symmetry broken WSM with a strong
low-frequency driving pulse of linear polarization. We have numerically
solved the semiconductor Bloch equations governing a WSM driven by a strong
laser field in the length gauge and considered the HHG process depending on
the Weyl node's momentum separation vector and the driving wave intensity.
Our results show that at the non-collinear arrangement of the electric and
Weyl node's momentum separation vectors, the anomalous harmonics are
generated which are polarized perpendicular to the direction of the pump
wave electric field. The intensities of anomalous harmonics are
quadratically dependent on the momentum space separation of the Weyl nodes.
When the right and the left Weyl fermions are merged, the anomalous
harmonics vanish. In contrast to the anomalous harmonics, the intensities of
normal harmonics do not depend on the Weyl node's momentum separation
vector. The dependences of the intensities of the normal and anomalous
harmonics on the intensity of the driving wave are completely different, and
for the moderately strong driving waves one can enter an extreme nonlinear
regime of HHG. The results of the current investigation are not only of
theoretical and academic importance but also will have significant
implications for the rapidly developing area of modern extreme nonlinear
optics of topological nanomaterials.

\begin{acknowledgments}
The work was supported by the Science Committee of Republic of
Armenia, project No. 21AG-1C014.
\end{acknowledgments}

\end{document}